\documentclass[10pt,sigconf,letterpaper,nonacm]{acmart}

\usepackage{amsmath}

\usepackage{amssymb,amsfonts}
\usepackage{array}
\usepackage{textcomp}
\usepackage{xcolor}
\usepackage{url}
\usepackage{listings}
\usepackage{algorithm}
\usepackage{algorithmic}
\usepackage{graphicx}
\usepackage{subfigure} 
\usepackage{fancybox}
\usepackage{xspace}
\usepackage{threeparttable}
\usepackage{multirow}
\usepackage{siunitx}
\usepackage{booktabs}
\usepackage{enumitem}

\renewcommand\footnotetextcopyrightpermission[1]{}

\begin{document}
\title{Seeing is (Not) Believing: Practical Phishing Attacks Targeting Social Media Sharing Cards} 

\author[W Huang]{Wangchenlu Huang}
\email{huangwang2019@bupt.edu.cn}
\affiliation{%
  \institution{Beijing University of Posts and Telecommunications}
  \city{Beijing}
  \country{China}
}

\author[S Wang]{Shenao Wang}
\email{shenaowang@hust.edu.cn}
\affiliation{%
  \institution{Huazhong University of Science and Technology}
  \city{Wuhan}
  \country{China}
}

\author[Y Zhao]{Yanjie Zhao}
\email{yanjie_zhao@hust.edu.cn}
\affiliation{%
  \institution{Huazhong University of Science and Technology}
  \city{Wuhan}
  \country{China}
}

\author[G Xu]{Guosheng Xu}
\email{guoshengxu@bupt.edu.cn}
\affiliation{%
  \institution{Key Laboratory of Trustworthy Distributed Computing and Service (MoE), Beijing University of Posts and Telecommunications}
  \city{Beijing}
  \country{China}
}

\author[H Wang]{Haoyu Wang}
\email{haoyuwang@hust.edu.cn}
\affiliation{%
  \institution{Huazhong University of Science and Technology}
  \city{Wuhan}
  \country{China}
}

\begin{abstract}
In the digital era, Online Social Networks~(OSNs) play a crucial role in information dissemination, with sharing cards for link previews emerging as a key feature. These cards offer snapshots of shared content, including titles, descriptions, and images. In this study, we investigate the construction and dissemination mechanisms of these cards, focusing on two primary server-side generation methods based on Share-SDK and HTML meta tags. Our investigation reveals a novel type of attack, i.e., Sharing Card Forgery~(SCF) attack that can be exploited to create forged benign sharing cards for malicious links. We demonstrate the feasibility of these attacks through practical implementations and evaluate their effectiveness across 13 various online social networks. Our findings indicate a significant risk, as the deceptive cards can evade detection and persist on social platforms, thus posing a substantial threat to user security. We also delve into countermeasures and discuss the challenges in effectively mitigating these types of attacks. This study not only sheds light on a novel phishing technique but also calls for heightened awareness and improved defensive strategies in the OSN ecosystem.
\end{abstract}

\maketitle

\section{Introduction}

Online Social Networks~(OSNs) have become integral to the modern digital communication landscape, predominantly due to their capacity to facilitate swift and effortless content sharing~\cite{alan2007measurement,eytan2012role}. Within this context, sharing cards for link previews~\cite{hassanexploratory,aviad2020chameleon,stivala2020deceptive} has emerged as a fundamental feature, offering a glimpse into the content being shared without clicking. These cards typically include a title, a brief description, and an image, collectively serving as a visual and informative representation of the linked content~\cite{twittercard}. By presenting a condensed yet informative preview, users can quickly assess the relevance and potential value of the shared content, thereby facilitating more informed decisions about whether to explore the linked material further. 

Despite the undeniable prevalence and influence of sharing cards in shaping user interactions on OSNs, their security implications have received scant attention from the research community. A recent study~\cite{li2022fakeshare} has revealed a concerning vulnerability, where the displayed source of a sharing card can be forged when generating the card through SDK invocations within applications. However, this research has adopted a narrow perspective, focusing solely on sharing cards within mobile applications while \textbf{overlooking the broader landscape of web-based sharing cards}.

In this paper, we introduce a novel type of attack, i.e., \textbf{Sharing Card Forgery attack (abbreviated SCF attack)}.
Specifically, this paper explores two primary mechanisms utilized in the generation of these sharing cards: Software Development Kits for Sharing~(Share-SDKs~\cite{jssdk,qqsdk,douyinsdk}) and HTML meta tags. Two of the most popular meta tag languages are Open Graph~(OG) protocol~\cite{OpenGraphProtocol} by Facebook and Twitter Cards~\cite{twittercard} by Twitter. While these mechanisms are efficient in delivering a streamlined user experience, they still have design flaws that can be exploited by attackers. By identifying these design flaws in existing sharing card implementations, we demonstrate two distinct attack vectors: \textit{short link redirection} and \textit{deceptive metadata injection}.

In \textit{short link redirection} attacks, a malicious actor can craft the title, description, and preview image for a shared card under the guise of an innocuous domain facade. They host the URL shortening service itself on this initially benign domain. When users click the shared shortened URL, it seamlessly redirects through the innocuous domain before reaching the ultimate malicious destination hosting malware, phishing sites, or other malicious content. This indirection technique allows the attacker to bypass the initial URL check performed by the Share-SDKs. In \textit{deceptive metadata injection} attacks, an attacker injects specially crafted HTML meta tags containing forged benign titles, descriptions, and preview images into the HTML of these malicious pages. The crux of the attack lies in divergent redirection behavior. When an crawler requests the page to scrape its metadata, it is redirected to a benign page with innocuous content matching the forged HTML meta tags. However, when an unwitting user visits the same URL, they are redirected to the malicious hosted content.

Through our analysis, we have underscored the feasibility of these vulnerabilities and demonstrated their potential impact across various OSNs. Regarding these vulnerabilities, our paper provides comprehensive insights and recommendations aimed at enhancing the security and reliability of sharing card mechanisms within OSNs. By addressing these issues, we can contribute to fostering a more secure online environment. Ultimately, our work highlights the importance of rigorous security assessments and the adoption of robust countermeasures to mitigate the risks posed by SCF attacks, thereby safeguarding users from falling victim to malicious actors exploiting these vulnerabilities.

In summary, this paper makes the following contributions:

\begin{itemize}[leftmargin=15pt]
    \item \textbf{Novel Attack.} We introduce a novel type of attack, i.e., SCF attack, targeting web applications, and make an in-depth exploration of two primary attack vectors called \textit{short link redirection} and \textit{deceptive metadata injection}.
    \item \textbf{Comprehensive Evaluation.} We conduct a comprehensive evaluation of the feasibility and potential impact of SCF attacks across six major OSNs. Through practical demonstrations and case studies, we underscore the severity and far-reaching consequences of these attacks, highlighting the urgent need for robust countermeasures.
    \item \textbf{Effective Mitigation.} We propose effective mitigation strategies and best practices to address the identified flaws, including guidelines for secure URL handling, metadata validation, and user protection mechanisms. These recommendations aim to empower platform developers and security professionals to strengthen their defenses against SCF attacks and similar exploitation tactics.
\end{itemize}
\section{Content Sharing Card}
\label{sec:Content Sharing Card}

As one of the most dominant forms of Internet content at present, the fundamental idea of content-sharing cards is to offer a pre-determined and straightforward graphical presentation of a URL link~\cite{stivala2020deceptive}. Therefore, some people also refer to these sharing cards as ``unfurl'' and the process of generating these cards as ``unfurling''~\cite{teamsunfurling}, which means ``to spread out from a rolled or folded state''. In terms of specific implementation details, we primarily focus on two different content-sharing card generation strategies: i) cards generated based on platform SDKs, and ii) cards generated based on HTML meta tags. 

\subsection{Content Sharing Based on Share-SDK}
\begin{figure}
    \centering
    \includegraphics[width=\linewidth]{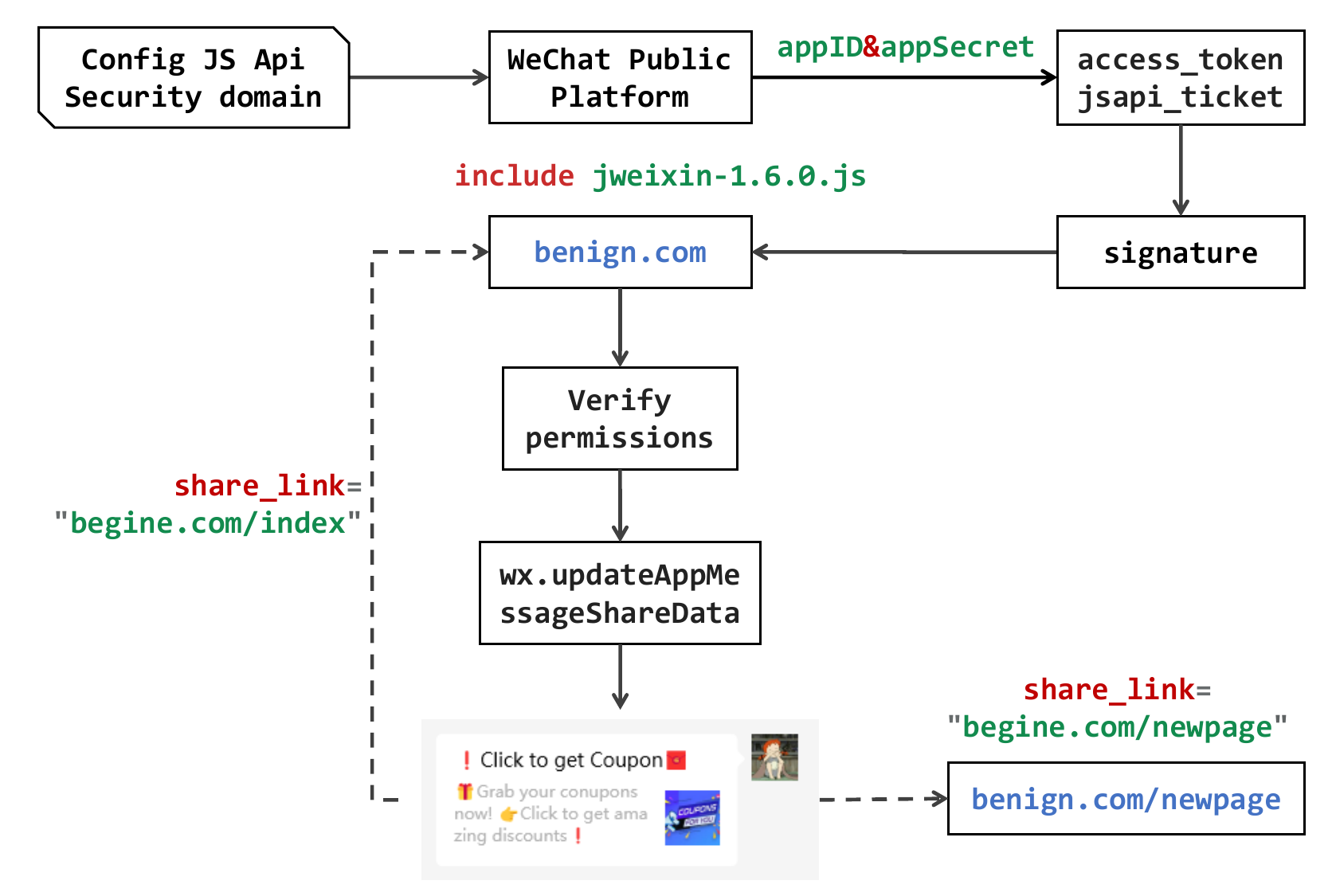}
    \caption{Sharing card generation based on Share-SDK.}
    \label{fig:SDK-card}
\end{figure}

Some OSNs choose to integrate the functionality of generating content-sharing cards into their SDKs, for example, WeChat~\cite{jssdk}, Douyin~\cite{douyinsdk}, QQ~\cite{qqsdk}, etc. Developers need to call the APIs provided in the SDK within their web pages. When users engage in sharing behavior, these functions are invoked to generate an attractive sharing card. As illustrated in Figure~\ref{fig:SDK-card}, we take WeChat as an example to elucidate the entire process of card generation.

\noindent\textit{\textbf{S1: Register the Domain.}}
Configure the ``JS API Secure Domain Name'' on the Weixin official accounts platform~\cite{jssdkconfig}. Once set up, developers can access WeChat's open JS interfaces through this domain name. The \texttt{access token} serves as the public number's globally unique interface invocation credentials. When accessing each interface, the web page must use the \texttt{access token}. Developers can use the \texttt{AppID} and \texttt{AppSecret} to obtain the \texttt{access token} by sending API requests~\cite{jssdkak}.

\lstset{
basicstyle=\normalsize\ttfamily,
numbers=left,
basicstyle=\tt,
numberstyle=\tiny\color{gray},
showstringspaces=false,        			
frame=single,                         	
}


\noindent\textit{\textbf{S2: Import JS Files and Validate Permissions.}}
The developers should import the WeChat-provided JS-SDK file into the webpage that requires access to the JS interface. 
Additionally, To leverage WeChat's JS-SDK APIs, developers must go through an authentication process enforced by the WeChat platform, by invoking \texttt{wx.config}. 

\noindent\textit{\textbf{S3: Call the Required API.}}
Then developers call the interface of \texttt{wx.updateAppMessageShareData} to customize the data display in the card~\cite{jssdk}. This function enables developers to modify the sharing card's title, description, and icon. They can also set jump links for shared actions, directing users to the original or a new page within the same domain.


\noindent\textit{\textbf{S4: User Access and Forwarding URL.}}
WeChat users can trigger function calls to form cards by accessing and forwarding URLs that have undergone the above operations.
During this sharing process, the link pointed to by the card is no longer displayed to the user as a URL but is fully forwarded and shared as a card. This is how traditional URLs are transformed into cards within WeChat.

\subsection{Content Sharing Based on HTML Meta Tags}
\begin{figure}[!htbp]
    \centering
    \includegraphics[width=0.7\linewidth]{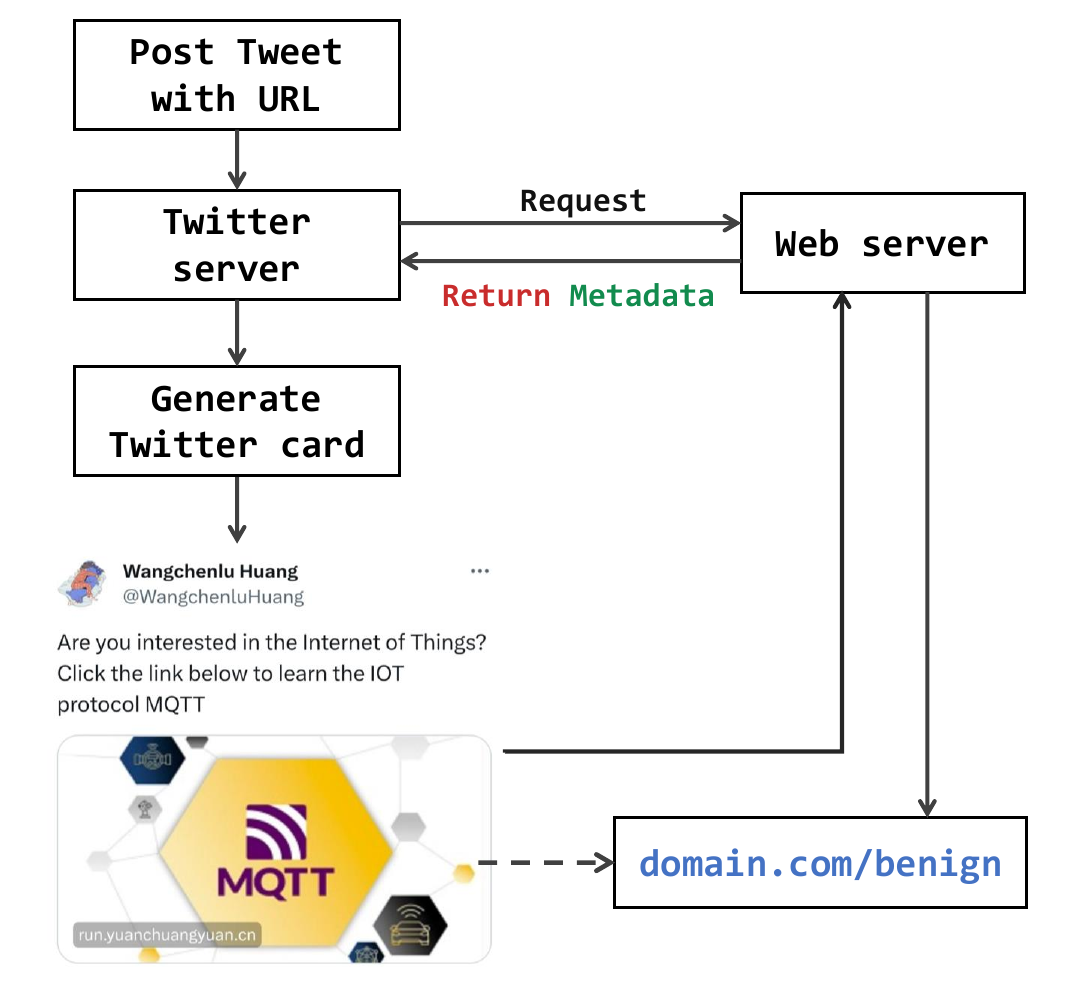}
    \caption{Sharing card generation based on OG protocol}
    \label{fig:web-card}
\end{figure}

An alternative method for creating sharing cards is through HTML meta tags. When a URL is uploaded into the OSNs using this method, the web crawler accumulates, stores, and displays the relevant metadata about the application or website. Most OSNs adhere to the standard metadata tags of the OG protocol, with some including their own extensions to customize their ecosystems. Twitter supports the proprietary format called ``Twitter Cards'' to enhance its functionality~\cite{twittercard} in its platform. It utilizes \texttt{<meta>} tags, but with specific Twitter attributes like \texttt{twitter:card}, \texttt{twitter:title}, \texttt{twitter:description}, and others. By doing so, when a user shares a link, a specialized crawler programmed in the website's structure is used by Twitter to examine the HTML data of the shared app or website. This process is exemplified by the following illustration in Figure~\ref{fig:web-card}. 

The Twitter server makes an HTTP request for the URL contained in the tweet, with the goal of obtaining metadata about the page~\cite{twittercard}. Twitter's crawler uses the User-Agent name Twitterbot. The version number~(like Twitterbot/1.0) might also be included. The target web server then processes the request and sends back an HTTP response containing metadata about the web page, including messages related to the card property in the HTML \texttt{<meta>} tag. The Twitter server analyses metadata to develop HTML and CSS codes for displaying cards in tweets. This guarantees proper depiction of cards across different clients and devices on Twitter. When a card is generated, it gets readily displayed on Twitter, along with the URL of the original share. For quick display when the same URL is shared, Twitter may cache this card information. If the page's content undergoes changes, the webmaster may need to request that Twitter recrawl the URL to update the information on the card.
\section{SCF Attack}
\label{sec:implementation}

In this section, we introduce SCF attacks. In response to the design flaws of Share-SDK and HTML meta tags, we propose two attack vectors, namely \textit{short link redirection} and \textit{deceptive metadata injection}.

\subsection{Design Flaws}
While in principle the content presented on social media link cards should closely relate to the website being shared, in practice this link is often tenuous or non-existent. An analysis of common card generation methods reveals two key issues that decouple the card content from the website content. The first issue arises with Share-SDKs provided by platforms like WeChat. These allow developers to hard-code the card properties like title, description, and imagery into the app's code, independent of the website's actual content. For example, WeChat's wx.updateAppMessageShareData() function takes these properties as parameters that can be set arbitrarily. This means the card can display content completely unrelated to what is found on the website itself. The second issue stems from the HTML meta tags, where card properties are specified via a set of <meta> tags in the HTML header. While visible to web crawlers, these tags are separate from the markup that renders the website visually for users. There is no inherent connection between the content in the meta tags and the actual page content below. This frees developers to design their link cards without being bound to represent the website accurately. As a result of these practices, social media link cards may bear little resemblance to the websites they putatively represent. Obviously, it runs counter to the principle that linked content should faithfully capture the underlying source.

\subsection{Attack Details}
In response to the two identified design flaws, we similarly propose two different ways to implement the sharing card forgery attack.

\begin{figure}[t]
    \centering
    \includegraphics[width=\linewidth]{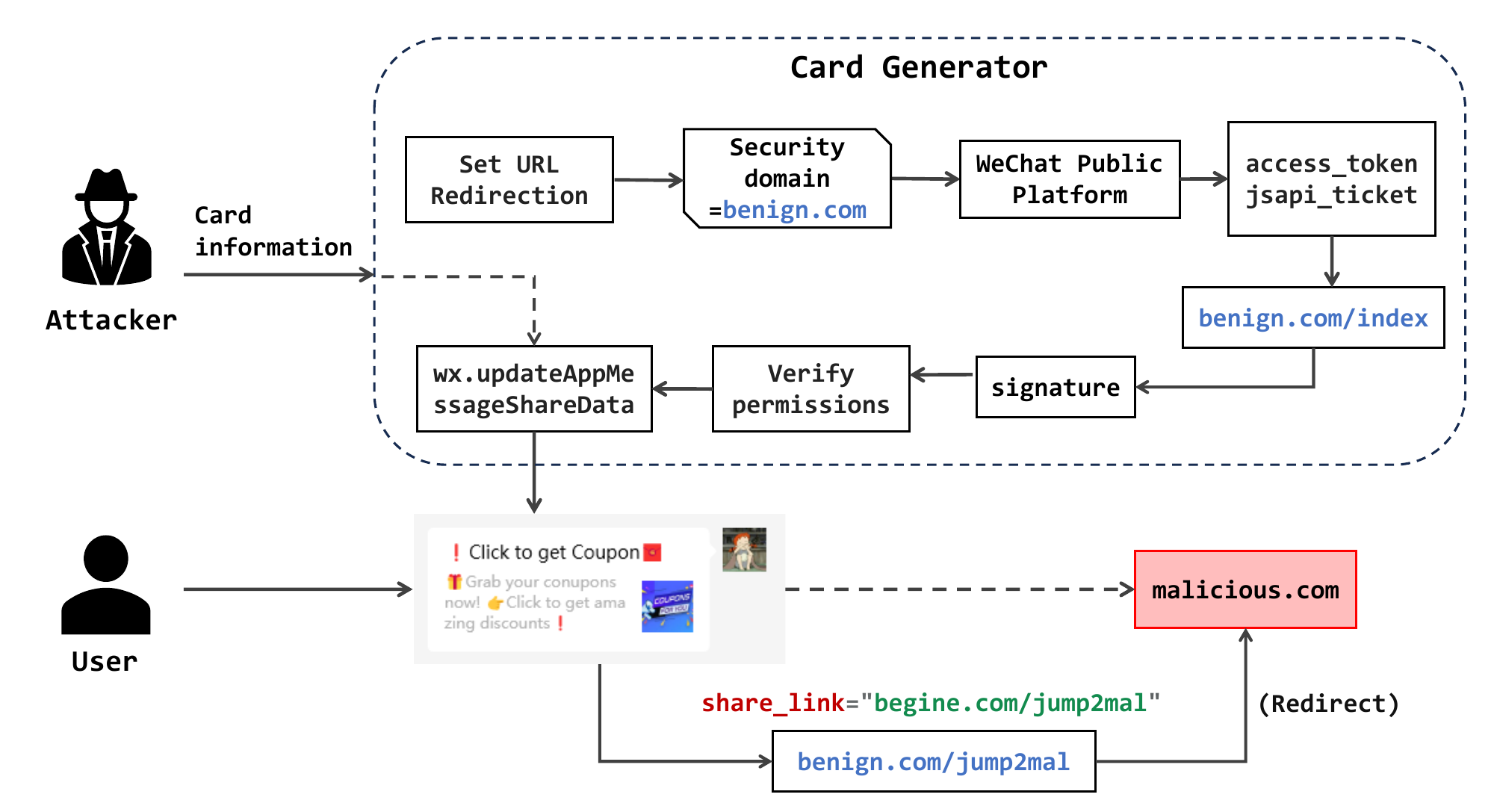}
    \caption{Short link redirection attack details}
    \label{fig:SDK-attack}
\end{figure}

\subsubsection{Short Link Redirection.}

The attacker first prepares the malicious URL they want to spread and then wraps that malicious URL through a short link service, generating a new benign URL like \texttt{benign.com/jump2mal} to serve as an intermediary step. This intermediary URL belongs to the benign domain, which can be configured in WeChat's developer platform and embed code containing the intermediary URL into a webpage~(\texttt{benign.com/index}) that imports WeChat's SDK. Attackers generate shared cards by embedding SDK reference code under benign domain names. Instead of pointing directly to the attacker's malicious site, this card links to the innocuous-looking intermediary URL~(\texttt{benign.com/jump2mal}). Users who click on the shared card are initially redirected to the intermediary URL. But behind the scenes, the intermediary URL instantly redirects the user's browser to the true malicious URL specified by the attacker, using redirection rules configured on the server. In essence, the short link redirection adds an extra hop that obscures the real malicious destination from the user. The innocuous intermediary URL makes the shared content appear safe, even though it automatically switches to the malicious site in the background. This redirection happens seamlessly without requiring any user interaction beyond the initial click.

\subsubsection{Deceptive Metadata Injection}

\begin{figure}
    \centering
    \includegraphics[width=0.8\linewidth]{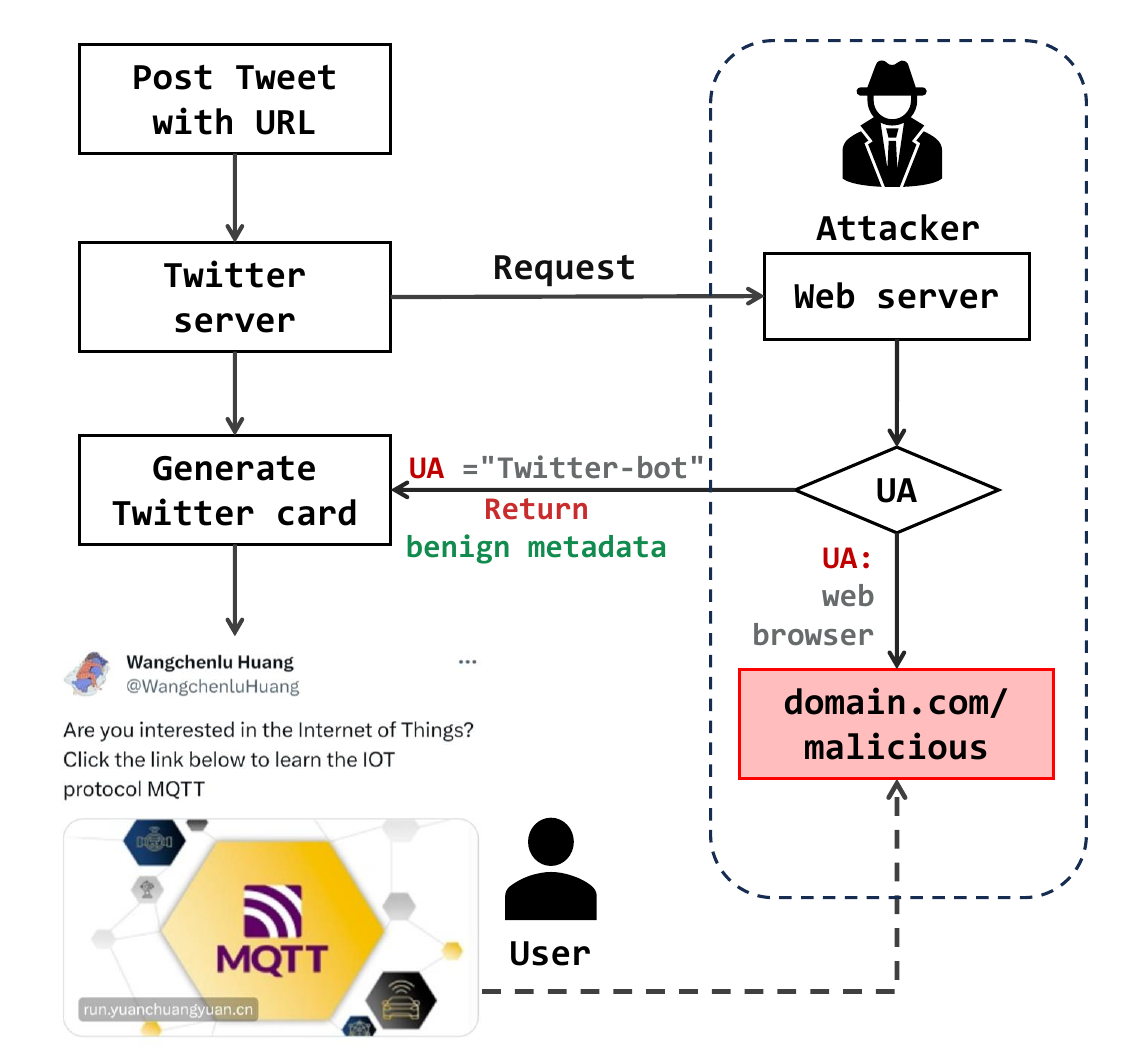}
    \caption{Deceptive metadata injection attack details}
    \label{fig:web-attack}
\end{figure}

For cards generated based on the HTML meta tags, the attacker does not need to resort to additional tools but only needs to add some functionality to the code implemented on the server side that contains the malicious web page, in order to share the card's message forgery. To describe the process, we take Twitter as an example, as shown in Figure~\ref{fig:web-attack}. The attacker posts a tweet containing a malicious URL. When the Twitter server receives a request for a URL, it generates a Twitter card, a process that involves the Twitter server sending a request to the website that provided the URL. The User-Agent in this request is typically set to ``Twitter-bot''. If the target site recognizes the request as coming from a crawler, it returns benign metadata to generate the Twitter card included in the tweet. When a normal user views the tweet and is tricked into clicking on the card, the attacker's server redirects the user to a malicious web page~(\texttt{domain.com/malicious}) because their user agent is usually a normal web browser identifier that does not contain the ``Twitter-bot''. In this way, the attacker can take advantage of the user's trust in Twitter and the benign cards to trick the user into clicking a malicious website.

\subsection{Potential Risks}
Reports~\cite{socialmediaphishing1,socialmediaphishing2} indicate that by 2021, more than 3.96 billion people worldwide will be using social media. On average, each social media user has 8.6 accounts on different platforms. This vast and diverse traffic provides endless opportunities for phishing threat actors. Unlike traditional email phishing attacks~\cite{alkhalil2021phishing}, SCF attacks occur in a dynamic and open social media environment. The appearance of the sharing card aligns with real platform content, making it more believable and harder to detect, thus increasing the success rate of phishing attacks. Its rapid dissemination on social media enables phishing links to reach a wider audience with minimal effort. By analyzing users' interests and networks, attackers can tailor sharing cards to specific interests, enhancing the relevance and effectiveness of phishing attacks.
\section{Evaluation}
\label{sec:evaluation}

Our evaluation is driven by the following three research questions (RQs):

\begin{enumerate}[label=\textbf{RQ\arabic*},labelsep=0.5em]
    \item \textit{How feasible is it to perform SCF attacks?}
    \item \textit{What OSNs are susceptible to SCF attacks?}
    \item \textit{How long is the alive period of the forgery card?}
\end{enumerate}

\subsection{Evaluation Setup}
To evaluate the severity and real-world impacts of the attacks described, we perform a two-pronged analysis approach. For the \textit{deceptive metadata injection} attack, we conduct a qualitative analysis examining the potential security implications. For the \textit{short link redirection} attack that facilitated the spreading of malicious links under a benign guise, we analyzed popular card generation tools and platforms actively in use. Specifically, we selected five mainstream tools that provide card/link creation interfaces as listed in Table~\ref{tab:card generator}

\subsection{RQ1: How feasible is it to perform SCF attacks?}
For OSNs that utilise HTML meta tags and web crawlers to generate sharecards, if an attacker is able to obtain the user agent corresponding to the crawler employed by the OSN and employs the aforementioned attack method, it is possible to modify the user agent of the HTTP request for authentication and detection purposes, thereby separating the content of the meta tags from the content displayed on the website. In theory, all OSNs that support meta tags are susceptible to successful attacks. The SCF attack was successfully executed on all seven OSNs that were subjected to testing. Given that HTML meta tags are currently an important tool for enterprises and marketers to optimise their content for social media sharing and increase engagement, their main application is to promote a specific brand or business on social media. The card must be designed in accordance with the specific requirements of the business in question. Consequently, a number of OSNs have not yet implemented validation mechanisms for auto-scraped metadata, as this would present significant challenges in terms of standardisation and design. This ultimately results in the ease with which our attack can be executed.

\begin{table}[t]
    \centering
    \caption{Sharing card forgery tools in the wild}
    \label{tab:card generator}
    \footnotesize
    \begin{tabular}{>{\arraybackslash}p{3cm}>{\centering\arraybackslash}p{3cm}>{\centering\arraybackslash}p{1.2cm}}
        \toprule
        \multicolumn{1}{m{3cm}}{Online Card Generator}
        & \multicolumn{1}{m{3cm}}{\centering Card Content Irregularity Detection Approach} & \multicolumn{1}{m{1.2cm}}{\centering Detecting URL Security?} \\
        \midrule
        Xiaohu Card~\cite{xiaohu} & Internal self-detection &  \\
        Tiantianwailian~\cite{tiantianwailian} & Internal self-detection &  \\
        AiDuanLian~\cite{aiduanlian} & Internal self-detection &  \\
        SanWeiTui~\cite{sanweitui} & Internal self-detection & \checkmark \\
        Xiaomark~\cite{xiaomark} & Internal self-detection &  \\
        WeiDuanLian~\cite{weiduanlian} & Internal self-detection & \checkmark \\
        Custom Sharing site~\cite{customsharing} & Not have &  \\
        Card Making Tool~\cite{cardmakingtool} & Not have &  \\
        \bottomrule
    \end{tabular}
\end{table}

In OSNs that rely on the Share-SDK to create shared cards, we researched and tested commonly used tools that invoke the SDK to generate shared cards, and found that most of them can be used as vectors for SFC attacks, and only a few of them are equipped with mechanisms for detecting malicious URLs, which can be broadly categorised into two main groups: those that perform security checks on input URLs, and those that focus on the detection of card content.

\subsubsection{Perform Security Checks on Input URLs}
According to our research and testing results, We found that in the process of generating cards, the vast majority of tools do not perform security checks on the URLs entered by the attackers. Most tools do not have URL blacklists or their blacklists cannot be kept shared and synchronized with social platforms, so they can share URLs that have been blocked or intercepted by social media. including domains that are not filed for recordation, web pages with malicious content, or web pages violating the norms of sharing on social platforms, etc.In our tests, there were only two tools that check the security of the input URL before creating a card. WeiDuanLian checks if the user's URL has an MIIT filing and if it doesn't, the card can't be generated. The second is the SanWeiTui Promotional tool, which does not restrict user input, but still evaluates the security of the URL. After a user clicks on a card that may be risky, the card will explicitly alert the user that the content they are about to access may be risky before the card completes the URL jump. Whether or not the user is alerted is most likely related to its internal whitelisting mechanism.

\subsubsection{Card Content Detection}
Some card generation tools have integrated improved content verification mechanisms to check for the presence of manipulative content in order to combat the problem. When the card cover is found to contain manipulative content, these verification functions, which are usually integrated within the tools themselves, take action to stop the redirection of short links. However, a number of card-generation tools are still in use without any kind of verification system. Only after user reports are the sharing cards generated by these tools blocked or banned. Moreover, it should be noted that even those tools that do have verification capabilities are limited to standardizing the text and image content on the cards; they have not been able to address the disparity or mismatch between the content of the card and the content of the URL link that is associated with it.

\subsection{RQ2: What OSNs are susceptible to SCF attacks?}
As we have highlighted earlier, our research indicates that the majority of OSNs currently supporting the HTML meta tags are susceptible to successful attacks. This includes major platforms such as Facebook~\cite{facebook}, Pinterest~\cite{pinterest}, Telegram~\cite{telegram}, Slack~\cite{slack}, Discord~\cite{discord}, WhatsApp~\cite{whatsapp}, and LinkedIn~\cite{linkedin}. These platforms, widely used for their extensive reach and effectiveness in disseminating information, become prime targets due to their reliance on the meta tags for rendering shared content. In the context of OSNs that generate sharing cards based on Share-SDK, the card generation tools play a decisive role. We conducted a survey and testing of commonly used and widely popular card generation tools currently available on the market. These tools integrate the SDKs of various social networking sites and social media platforms, as shown in Table~\ref{tab:platforms}, covering WeChat~\cite{wechat}, TikTok~\cite{tiktok}, QQ~\cite{qq}, Zhihu~\cite{zhihu}, WeCom~\cite{wecom}, Bilibili~\cite{bilibili}. This expands the range of phishing attacks available to perpetrators, subjecting victims to a broader and more complex set of phishing threats. It is evident that WeChat, as the application most frequently used for daily communication by the public, is the primary target application of card generation tools.

\begin{table}[!htbp]
    \caption{OSNs platforms that are vulnerable to SCF attack}
    \label{tab:platforms}
    \footnotesize
    \begin{tabular}{>
    {\arraybackslash}p{2.4cm}>
    {\centering\arraybackslash}p{0.6cm}>
    {\centering\arraybackslash}p{0.6cm}>
    {\centering\arraybackslash}p{0.6cm}>
    {\centering\arraybackslash}p{0.5cm}>
    {\centering\arraybackslash}p{0.5cm}>
    {\centering\arraybackslash}p{0.6cm}}
        \toprule
        \multicolumn{1}{m{2.4cm}}{Card Generator}
        & \multicolumn{1}{m{0.6cm}}{\centering WeChat} 
        & \multicolumn{1}{m{0.6cm}}{\centering TikTok}
        & \multicolumn{1}{m{0.5cm}}{\centering QQ}
        & \multicolumn{1}{m{0.5cm}}{\centering Zhihu}
        & \multicolumn{1}{m{0.6cm}}{\centering WeCom}
        & \multicolumn{1}{m{0.6cm}}{\centering Bilibili} \\
        \midrule
        Xiaohu Card~\cite{xiaohu}                 & \checkmark & \checkmark & \checkmark & \checkmark & \checkmark & \\
        TianTianWaiLian~\cite{tiantianwailian}             & \checkmark & \checkmark &            &            & \checkmark & \\
        AiDuanLian~\cite{aiduanlian}            & \checkmark & \checkmark &            & \checkmark &            & \\
        SanWeiTui~\cite{sanweitui}  & \checkmark & \checkmark \\
        Xiaomark~\cite{xiaomark}      & \checkmark & & & & \checkmark \\
        WeiDuanLian~\cite{weiduanlian}           & \checkmark & \checkmark & \checkmark & & & \checkmark \\
        Custom Sharing~\cite{customsharing}         & \checkmark &  \\
        Card Making Tool~\cite{cardmakingtool}            & \checkmark &  \\
        \bottomrule
    \end{tabular}
\end{table}

\subsection{RQ3: How long is the alive period of the forgery card?}
In the context of evaluating phishing websites, the duration of their operation (from the time they go live to when they are shut down) is particularly critical. Firstly, the longer a phishing site remains active, the greater the opportunity to deceive victims, leading to the leak of sensitive information. Through experimental testing of the two attack methodologies discussed above, we have observed the following: In the scenario where the attack leverages the User-Agent to induce the creation of false cards, the stability of these fabricated cards is notable. The presence of misleading content within these cards does not lead to their banishment by the platform. Consequently, the cards can be viewed and accessed until they are mass redistributed or reported by users. In attacks involving card generation tools, some tools limit user input, preventing attackers from creating false cards. However, some of these tools examine the card content, resulting in varying survival durations for falsely generated cards depending on the tool used. Tools with verification mechanisms do not restrict card creation or user access; instead, they perform semantic matching checks or manual inspections of cards after they have been created. Cards with overtly rule-breaking, misleading content on the AiDuanLian platform typically last about 10 hours if they are not widely disseminated. Furthermore, the platform has the right to completely ban an account if it is discovered that the user has created two such cards in violation. Cards have a two-day survival duration on the TianTianWaiLian and XiaoHu platforms. Cards on the Xiaomark platform frequently last longer than three days. False cards may persist in tools without verification mechanisms until users report them or the OSNs identify them as problematic. This analysis emphasizes the need for strong verification procedures to stop the spread of harmful or misleading content on social media platforms, as well as the varying efficacy of various card generation tools in mitigating such attacks.
\section{Mitigation}
The attacks explored in this work highlight critical vulnerabilities in how social media cards and URLs are generated and rendered. To mitigate these risks, steps need to be taken by platforms, developers, and users.

\noindent \textbf{Platform-Level Mitigations.}
Social media platforms should implement stricter validation checks on the card metadata and linked URLs accepted from third-party apps and websites. Rather than blindly trusting provided metadata, platforms could perform content audits that inspect the underlying website for consistency with the specified title, description, imagery, etc. 
For short links, platforms should analyze the redirection chains and apply reputation checks on the final destination URLs, blocking known malicious sites. Requiring direct links without redirections could also mitigate some redirection-based attacks.

\noindent\textbf{Developer Mitigations.}
App and website developers should implement robust systems that generate social media cards dynamically based on the core site content, rather than allowing arbitrary metadata specification.
Additionally, developers should avoid client-side fabricated links and redirections. Server-side link generation with strict validation checks on inputs can prevent injecting malicious destinations.
\section{Related Work}
Several recent studies have investigated the security implications of link previews on popular social media platforms. Stivala et al.~\cite{stivala2020deceptive} conducted a comprehensive study analyzing the link preview rendering process across 20 social networks and instant messaging apps. They discovered that four platforms~(Facebook, Xing, Plurk, and Slack) allowed an attacker to craft benign-looking previews for malicious links, potentially misleading even tech-savvy users. Elyashar et al.~\cite{aviad2020chameleon} introduced ``chameleon'' attacks, where an attacker can manipulate the content display on multiple social media platforms by creating a mismatch between the malicious page content and its benign-looking link preview. Kaleli et al.~\cite{kaleli2021unintended} studied unintended URL sharing on social media, showing how user mistakes or adversary actions can lead to unintentionally sharing sensitive information, malicious links, or explicit content. More recently, Xu et al.~\cite{li2022fakeshare} have revealed the Fake-Share attack in Android applications, where the displayed source of a sharing card can be forged when generating the card through SDK invocations. In contrast, our work provides a more comprehensive investigation into card generation mechanisms across OSNs and introduces two attack vectors, which reveal endemic risks in the shared card ecosystem that extend beyond individual platform vulnerabilities.
\section{conclusion}
In this paper, we introduced the SCF attack, a novel class of attacks targeting web applications through two primary vectors: short link redirection and deceptive metadata injection. Our evaluation across major OSNs demonstrated the feasibility and impact of these attacks in facilitating malware spread and phishing campaigns. To this end, we have proposed effective mitigation strategies and best practices. Our work serves as a call to action for the community to prioritize the security of sharing cards, ensuring that users can engage with shared content confidently and without falling prey to deceptive tactics.

\bibliographystyle{ACM-Reference-Format}
\bibliography{main}

\end{document}